\documentstyle[pre,aps,epsfig,multicol,psfig]{revtex}

\begin{document}
\draft

\title{Front speed enhancement in cellular flows}

\author{M. Abel$^{(1,2)}$, M. Cencini$^{(2,3)}$, D. Vergni$^{(2,3)}$ and
A. Vulpiani$^{(2,3)}$} 
\address{$^{(1)}$Institute of Physics, Potsdam
University, 14415 Potsdam, Germany} 
\address{$^{(2)}$ Dipartimento di
Fisica , Universit\`a di Roma ``La Sapienza'', \\
P.zzle Aldo Moro 2, I-00185 Roma, Italy } 
\address{$^{(3)}$ Istituto Nazionale Fisica della Materia
UdR and SMC Roma ``La Sapienza''} 
\maketitle

\begin{abstract}
The problem of front propagation in a stirred medium is addressed in
the case of cellular flows in three different regimes: slow reaction,
fast reaction and geometrical optics limit.  It is well known that a
consequence of stirring is the enhancement of front speed with respect
to the non-stirred case.  By means of numerical simulations and
theoretical arguments we describe the behavior of front speed as a
function of the stirring intensity, $U$. For slow reaction, the front
propagates with a speed proportional to $U^{1/4}$, conversely for fast
reaction the front speed is proportional to $U^{3/4}$. In the
geometrical optics limit, the front speed asymptotically behaves as
$U/\ln U$.

PACS:  
\end{abstract}

\begin{multicols}{2}


{\bf Front propagation in a stirred medium is an important problem in
a number of fields ranging from combustion to plankton dynamics.  For
a realistic study of such a class of problems one has to take into
account the modification of the advecting flow induced by the
reaction, e.g. in combustion. However, many features can be understood
by neglecting the back-reaction on the velocity field.  The problem
addressed here is the enhancement of the front speed induced by a
certain class of flows.  In particular, we consider front propagation
in a two dimensional laminar flow with a stationary vortical structure
in different regimes, namely slow reaction, fast reaction and
geometrical optics limit. This last limit corresponds to a very sharp
front propagating as an optical front, i.e., according to the Huygens
principle.  We provide predictions on the dependence of the front
speed on the flow intensity, which are confirmed by numerical
simulations.}

\section{Introduction}
\label{sec:1}
The study of front propagation of a stable phase into an unstable one
encompasses several issues of great interest \cite{xin} as flame
propagation in gases \cite{combustion}, population dynamics of
biological communities (plankton in oceans) \cite{bio} and chemical
reactions in liquids \cite{chem}.  A common feature of all these
phenomena is that they take place in a strongly deformable medium such
as a fluid.  The interplay among transport, diffusion and reaction is
therefore a crucial problem with several open issues (e.g. for
questions concerning combustion see Ref.~\cite{Ronn1}).

In the most compact model of front propagation the state of the system
is described by a single scalar field $\theta({\bf r},t)$, that
represents the concentration of products. The field $\theta$ vanishes
in the regions filled with fresh material (the unstable phase), equals
unity where only inert products are left (the stable phase) and takes
intermediate values wherever reactants and products coexist, i.e., in
the region where production takes place.  Here we assume that the
concentration of chemicals does not modify the underlying
flow. Therefore, in the following, we consider the velocity field as
given. This approximation, hardly tenable in the context of flame
propagation in gases, is rather appropriate for chemical front
propagation in some liquid solutions \cite{combustion,Ronn1,Peter,ronney}.
Under these simplifying assumptions, the evolution of $\theta$ is
described by
\begin{equation}
\partial_t \theta + {\bf u}\cdot{\boldmath{\mbox{$\nabla$}}}\theta = D \Delta \theta + 
{1 \over \tau}f(\theta) \;,
\label{eq:rad} 
\end{equation}
 where the second term on the l.h.s. accounts for the transport by an
incompressible velocity field.  On the r.h.s the first term describes
molecular diffusion and the second one describes the production
process with time scale $\tau$. We will first consider a production
term of Fischer-Kolmogorov-Petrovski-Piskunov \cite{FKPP37} (FKPP)
type, i.e., a function $f(\theta)$ convex ($f{''}(\theta)<0$) and
positive in the interval $(0,1)$, vanishing at its extremes, and
$f'(0)=1$.  Here we take $f(\theta)=\theta(1-\theta)$.  It is also of
interest to consider a production term in the form of the Arrhenius
law, $f(\theta)=(1-\theta)\cdot\exp(-\theta_c/\theta)$, where
$\theta_c$ is the activation concentration. The latter choice is more
pertinent to the study of flames and/or chemical reactions
\cite{Peter,ronney}.

Until now we did not specify any details of the velocity field.  In
many engineering applications ${\bf u}$ is turbulent. In this paper we
investigate front propagation in laminar flows, which, albeit simpler
than turbulent ones, show remarkable qualitative similarities with
more complex flows \cite{lagrangian}.  Specifically, we consider a two
dimensional stationary incompressible flow with cellular structure
(see also \cite{Pomeau,Const,Ryzhik}) ${\bf u}=(-\partial_y
\psi,\partial_x \psi)$ with the streamfunction \cite{Gollub}
\begin{equation}
   \psi(x,y)={UL \over 2\pi} \sin\left({2\pi x\over L}\right)
 \sin\left({2\pi y\over L}\right) \,.
   \label{eq:stream}
\end{equation}
We considered $L$-periodic boundary conditions in $y$ and an infinite
extent along the $x$-axis.  This kind of flow is interesting because,
in contrast to shear flows, all the streamlines are closed and,
therefore, the front propagation is determined by the mechanisms of
contamination of one cell to the other \cite{Pomeau,abeletal}.  Since
we are interested in the propagation in the $x$-direction, the
boundary conditions are set to $\theta(-\infty,y;t)=1$ and
$\theta(+\infty,y;t)=0$.  The maximum principle ensures that at later
times the field still takes values in the range $0 \le \theta \le 1$
\cite{xin}.  The instantaneous front speed is defined as
\begin{equation}
v_{\mbox{\scriptsize f}}(t)={1 \over L} \int_{0}^{L} {\rm d}y
\int_{-\infty}^{\infty}{\rm d}x \, \partial_t \theta(x,y;t)\,.
\label{eq:velocity}
\end{equation}
This expression defines the so-called bulk burning rate \cite{Const}
which coincides with the front speed when the latter exists, but it is
also a well defined quantity even when the front itself is not well
defined.  The asymptotic (average) front speed, $v_{\mbox{\scriptsize
f}}$, is determined by $v_{\mbox{\scriptsize f}}=\lim_{T\to \infty}
1/T \int {\rm d}t \, v_{\mbox{\scriptsize f}}(t)$.

In a medium at rest, it is known that Eq.~(\ref{eq:rad}), for FKPP
nonlinearity, generates a front propagating, e.g., from left to right
with an asymptotic speed $v_0 = 2\sqrt{D/\tau}$ and a reaction region
of thickness $\xi =8\sqrt{D \tau}$ \cite{FKPP37}.  In the more
interesting case of a moving medium, the front will propagate with an
average speed $v_{\mbox{\scriptsize f}}$ greater than $v_0$
\cite{Const,Ryzhik}.  The front velocity $v_{\mbox{\scriptsize f}}$ is
the result of the interplay among the flow characteristics
(i.e. intensity $U$ and length-scale $L$), the diffusivity $D$ and the
production time scale $\tau$.  The goal of our analysis is to
determine the dependence of $v_{\mbox{\scriptsize f}}$ on such
quantities. In particular, introducing the Damk\"ohler number
$Da=L/(U\tau)$ (the ratio of advective to reactive time scales) and
the P\'eclet number $Pe=UL/D$ (the ratio of diffusive to advective
time scales), we seek for an expression of the front speed as an
adimensional function $v_{\mbox{\scriptsize f}}/v_0 = \phi(Da,Pe) \ge
1$. We will see that a crucial role in determining such a function is
played by the renormalization of the diffusion constant and chemical
time scale induced by the advection \cite{abeletal,spagnoli}.

Moreover, we consider an important limit case, i.e., the so called
geometrical optics limit, which is realized for $(D, \tau) \to 0$
maintaining $D/\tau$ constant \cite{Majda-Geq}. In this limit one has
a non zero bare front speed, $v_0$, while the front thickness $\xi$
goes to zero, i.e., the front is sharp. In this regime the front
dynamics is well described by the so-called $G$-equation
\cite{combustion,Peter,KAW88}
\begin{equation}
{\partial G \over \partial t}+{\bf u}\cdot {\mbox{\boldmath $\nabla$}}
G=v_0 |{\mbox{\boldmath $\nabla$}}G|\,.
\label{eq:optics}
\end{equation}
The front is defined by a constant level surface of the scalar
function $G({\bf r},t)$. Physically speaking, this limit corresponds
to situations in which $\xi$ is very small compared with the
other length scales of the problem. Also in this case we provide a
simple prediction for the front speed, which turns out to be
expressible as an adimensional function $v_{\mbox{\scriptsize f}}/v_0
= \psi(U/v_0)$.

The paper is organized as follows. In Sect.~\ref{sec:2} we discuss a
theoretical upper bound for the front speed which becomes an equality
in the limit of (very) slow reaction. In Sect.~\ref{sec:3} we present
a numerical study for slow and fast reaction, comparing the results
with a phenomenological model.  In Sect.~\ref{sec:4} we consider the
geometrical optics limit. Sect.~\ref{sec:5} is devoted to some
concluding remarks. The Appendix contains the details of the numerical
method used in the simulations.

\section{Upper bound for the front speed}
\label{sec:2}
For a generic incompressible flow and a generic production term which
has a bounded growth rate, $c(\theta)$, i.e.,
\begin{equation}
c_{\mbox{\scriptsize max}}=\sup_{\theta} c(\theta) = 
\sup_{\theta} {1 \over \tau}{f(\theta) \over \theta} < \infty\,\,,
\label{eq:growthrate}
\end{equation}
it is possible to establish an upper bound for the speed of front
propagation.    Explicitly, we have
\begin{equation}
v_{\mbox{\scriptsize f}} \le 
2\sqrt{D_{\mbox{\scriptsize  eff}} \, c_{\mbox{\scriptsize  max}}}
\label{eq:upb}
\end{equation}
where $D_{\mbox{\scriptsize eff}}$ is the effective diffusion
coefficient in the $x$-direction, which can be derived from
Eq.~(\ref{eq:rad}) by switching off the production term.  This bound
is the consequence of the deeply rooted link existing between front
propagation and advective transport.

We start the derivation of Eq.~(\ref{eq:upb}) by recalling the fundamental
relation among the solution of the PDE (\ref{eq:rad}) and the
trajectories of particles advected by a velocity field ${\mathbf
u}({\mathbf r},t)$ and subject to molecular diffusion
\cite{freid,fedotov} 
\begin{equation}
\theta({\mathbf x}, t) = \left \langle {
\theta({\mathbf r}(0), 0)\, e^{\scriptsize {\,\int_0^t
c(\theta({\mathbf r}(s),s)) {\mathrm d}s}}} \right \rangle_\eta\,.
\label{eq:feykacext} 
\end{equation}
The average is performed over the
trajectories evolving according to the Langevin equation 
\begin{equation} 
{{\rm d} {\mathbf r}(t) \over {\rm d}t}=
{\mathbf u}({\mathbf r}(t),t)+\sqrt{2D}\, \mbox{\boldmath $\eta$}(t)
\label{eq:langevin}
\end{equation}
with ending point ${\mathbf r}(t) = {\mathbf x}$. The white noise
term $\sqrt{2D}\mbox{\boldmath $\eta$}(t)$ 
accounts for molecular diffusion.\\
Since the growth rate is bounded, (\ref{eq:feykacext})
yields the inequality 
\begin{equation}
\theta(t,{\mathbf x})
\le 
\left\langle \theta({\mathbf r}(0),0) \right\rangle_{\eta}  
\exp(c_{\mbox{\scriptsize max}}t)\;.
\label{eq:ineq}
\end{equation}
In the previous inequality, the term in angular brackets denotes the
probability that the trajectory ending at ${\mathbf x}$ was initially
located at the left of the front interface.  For FKPP production term
the maximum occurs for $\theta=0$, i.e., $c(\theta) \le
c_{\mbox{\scriptsize max}}=c(0)=1/\tau$.  Under very broad conditions,
i.e., nonzero molecular diffusivity and finite variance of the
velocity vector potential~\cite{MA89-AV95,K70,MK99}, it is possible to
show that asymptotically the particles undergo a standard diffusion
process with an effective diffusion coefficient $D_{\mbox{\scriptsize
eff}}$, always larger than the molecular value $D$. The issue of
single particle diffusion, and the problem of finding the effective
diffusivity has been the subject matter of a huge amount of work (see
e.g.~\cite{MK99} for a recent review).  In the presence of an
asymptotic standard diffusion, we can substitute the term
$\left\langle \theta({\mathbf r}(0),0) \right\rangle_{\eta}$, with the
Gaussian result $1-{1 \over 2}\mbox{erfc}(-x/\sqrt{2
D_{\mbox{\scriptsize eff}} \,t})\simeq \exp[-x^2/(4
D_{\mbox{\scriptsize eff}}\, t)] /\sqrt{2 \pi D_{\mbox{\scriptsize
eff}}\, t}$ with exponential accuracy.  We thus obtain
$\theta({\mathbf x},t) \le \exp\left[c_{\mbox{\scriptsize max}}t-
x^2/(4D_{\mbox{\scriptsize eff}}\,t)\right] /\sqrt{2 \pi
D_{\mbox{\scriptsize eff}}\, t}$.  It is thus clear that at the point
${\mathbf x}$ the field $\theta$ is exponentially small until a time
$t$ of the order of $x/\sqrt{4 D_{\mbox{\scriptsize eff}}\,
c_{\mbox{\scriptsize max}}}$.  We finally obtain the upper bound for
the front velocity $v_{f} \le \sqrt{4 D_{\mbox{\scriptsize eff}}\,
c_{\mbox{\scriptsize max}}}$, which is Eq.~(\ref{eq:upb}).

The bound~(\ref{eq:upb}) becomes an equality in the limit of very slow
reaction.  If $\tau$ is the slowest time scale under consideration,
advection and molecular diffusion act jointly to build an effective
diffusion process, essentially unaffected by the reaction.  In this
case the front width is large enough and the reaction takes place in a
region of effective diffusivity. Therefore, it is allowed to
substitute Eq.~(\ref{eq:rad}) with an effective reaction-diffusion
equation $\partial_t \theta = \sum_{i,j} D^{\mbox {\scriptsize
{eff}}}_{ij} \partial^2_{ij} \theta + {1 \over \tau} f(\theta)$, where
$D^{\mbox{\scriptsize {eff}}}_{ij}$ is the eddy-diffusivity tensor
\cite{MK99}.  For the FKPP nonlinearity, this last equation gives rise
to $v_{\mbox{\scriptsize f}} \simeq 2 \sqrt{D_{\mbox{\scriptsize
eff}}/\tau}$, where $D_{\mbox {\scriptsize eff}}= D_{11}^{\mbox
{\scriptsize eff}}$ \cite{Ronn1,Pomeau,Const}.  One can find a
detailed derivation of this formula in Ref.~\cite{abeletal}.

For cellular flows, it is known that $D_{\mbox{\scriptsize eff}}\sim
\sqrt{UL D}$ \cite{Pom,Shr,Ros}.  Inserting this expression in
Eq.~(\ref{eq:upb}) one obtains $v_{\mbox{\scriptsize f}} \propto
U^{1/4}$ \cite{Pomeau}, remarkably close to the observed ones for $Da
\ll 1$ (see next Section).

\section{Front speed in the Reaction Advection Diffusion equation}
\label{sec:3}

The bound~(\ref{eq:upb}) is very general and holds for generic
incompressible flows and production terms.  Here, by means of
numerical simulations, we consider the front propagation problem
arising in the reaction advection diffusion equation~(\ref{eq:rad})
for the particular case of the cellular flow (\ref{eq:stream}) of
stirring intensity $U$ and FKPP nonlinearity (with characteristic
time, $\tau$). In our discussion, we always suppose that the diffusion time
scale is the slowest occurring one, i.e., ${L^2 / D} \ll
{L / U},\tau$ and thus $Pe \gg 1$ and $Da \, Pe \ll 1$.

Now before presenting the numerical results it is helpful to introduce
a macroscopic description of the problem which will reduce it to an
effective reaction-diffusion process with renormalized coefficients.

The basic observation is that the dynamics of $\theta$ in cellular
flows is characterized by the the cell-size $L$, so that we can
perform a space discretization that reduces each cell, $C_i$, to a
point, $i$, mapping the domain -- a two-dimensional infinite strip --
onto a one-dimensional lattice.  The field $\theta$ becomes a function
defined on the lattice $\Theta_i = L^{-2}
\int_{C_i} \theta \,{\rm d}x\,{\rm d}y$.  Integrating
Eq.~(\ref{eq:rad}) over the cell $C_i$, we obtain $\dot{\Theta}_i =
J_{i+1}-J_{i} +\chi_i$ where $J_{i}=L^{-2} \int_{\mbox{\scriptsize
left}} D \partial_x\theta\,{\mathrm d}y$ is the flux of matter through
the left boundary of the $i$-th cell, and $\chi_i=L^{-2}\int_{C_i}
\tau^{-1} f(\theta) \,{\mathrm d}x\,{\mathrm d}y$ is the rate of
change of $\Theta_i$ due to reaction taking place within the cell.  On
the basis of the numerical results (see below), we will show that the
space-discretized macroscopic reaction-diffusion equation
\begin{equation}
{{\mathrm d} \over {\mathrm d}t} \Theta_i= 
D_{\mbox{\scriptsize eff}}\left(\frac{1}{2} \Theta_{i+1} -\Theta_i + 
{1 \over 2}\Theta_{i-1}\right) + 
{1 \over \tau_{\mbox{\scriptsize eff}}} F(\Theta_i)
\label{eq:3}
\end{equation}
is a pretty good model for the front dynamics.  The effect of the
advective field is to {\em renormalize} the values of the
diffusivity, $D \to D_{\mbox{\scriptsize eff}}(D,U,L)$, and the
reaction time scale, $\tau \to \tau_{\mbox{\scriptsize
eff}}(\tau,U,L)$.  This is why the velocity does not appear any more
in the effective dynamics, described by  Eq.~(\ref{eq:3}) 
\cite{Pomeau,abeletal}.
The renormalized diffusivity $D_{\mbox{\scriptsize eff}}$ accounts
for the process of diffusion from cell to cell as a result of the
nontrivial interaction of advection and molecular diffusion
\cite{Pom,Shr,Ros}. The renormalized reaction time
$\tau_{\mbox{\scriptsize eff}}$ amounts to the time that it takes for
a single cell to be filled by inert material, and depends on the
interaction of advection and production.  Of course, also the
production term will be affected, i.e.  $f \to F$, but, as we will
see, such a modification is not dramatic. Indeed, the modified
production term turns always to be in the FKPP universality class.

The limiting speed of the front in the moving medium turns
out to be $v_{\mbox{\scriptsize eff}}\sim \sqrt{D_{\mbox{\scriptsize
eff}}/\tau_{\mbox{\scriptsize eff}}}$, similar to Eq.(\ref{eq:upb})
\cite{abeletal}.  The problem is now reduced to derive the expressions
for the renormalized parameters by means of physical considerations.

In the following sections, using as an interpretative framework the above
described macroscopic model, we will present the results of detailed
numerical simulations for slow ($Da\ll 1$) and fast ($Da \gg 1$)
reaction. 

\subsection{Slow reaction regime}
\label{sec:3.2}

At small $Da$, the reaction is significantly slower than the advection,
and consequently
the region where the reaction takes place extends over several cells,
i.e., the front is distributed.  
To obtain the expression of $D_{\mbox{\scriptsize eff}}$ we neglect the
reaction term in Eq.~(\ref{eq:rad}), which reduces to the equation for 
a passive scalar in a cellular flow. 
This is a well studied problem, the solution of which is \cite{Pom,Shr,Ros}:
\begin{equation}
{D_{\mbox{\scriptsize eff}}\over D} \sim
Pe^{1/2} \qquad Pe \gg 1 \;.
\label{eq:4}
\end{equation}
For large $Pe$ ($D$ small) the cell-to-cell diffusion mechanism can be
qualitatively understood in the following way. The probability, $p$, for a
particle to jump across the boundary of the cell, within a circulation
time $L/U$, by virtue of molecular diffusion can be estimated as the
ratio of the diffusive motion across streamlines, $O(\sqrt{D L/U})$,
to advective motion along streamlines, $O(L)$. As a result
$p\sim(D/(UL))^{1/2}$, hence the effective diffusivity
$D_{\mbox{\scriptsize eff}}\sim p\,UL \sim D Pe^{1/2}$.

\begin{figure} 
\centerline{\includegraphics[scale=0.23,draft=false]{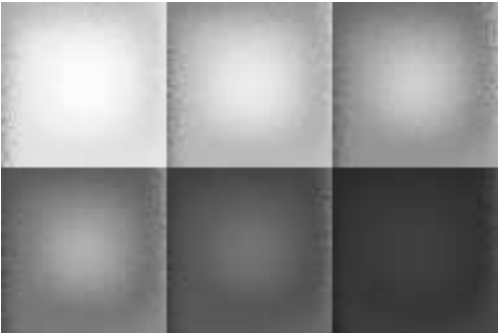}}
\vspace{0.4truecm}
\caption{Six snapshots of the field $\theta$ within the same cell,
at six successive times with a delay $\tau/6$ (from left to right, top
to bottom), as a result of
the numerical integration of equation~(\protect\ref{eq:rad}).
Here $Da \simeq 0.4,Pe\simeq 315$. Black stands for
$\theta=1$, white for $\theta=0$.} 
\label{fig:1}
\end{figure}

\begin{figure} 
\centerline{\includegraphics[scale=0.33,draft=false,angle=270]{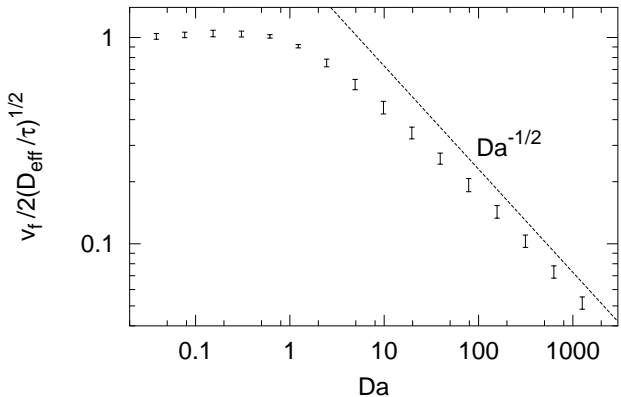}}
\caption{The ratio of measured front speed, $v_{\mbox{\scriptsize eff}}$, 
to the maximal one, $\sqrt{4 D_{\mbox{\scriptsize eff}}/\tau}$, 
as a function of the Damk\"ohler number, $Da$. For $Da \ll 1$ the 
front propagates with the maximal velocity  whereas
for $Da \gg 1$ the speed slows down with increasing $Da$. 
The behavior $Da^{-1/2}$ is drawn for reference.}
\label{fig:2}
\end{figure}

To obtain the expression of the typical time it takes for a whole cell
to react, let us concentrate on the reaction in a single cell: it is
first invaded by a mixture of reactants and products (with a low
content of products, $\Theta_i \ll 1$) on the fast advective
time scale; subsequently complete reaction ($\Theta_i\approx 1$) is 
achieved on the slower time scale $\tau_{\mbox{\scriptsize
eff}}\simeq\tau$ (see Fig.~\ref{fig:1}). In this regime, the front
speed is well approximated the homogenization result
$v_{\mbox {\scriptsize {f}}}=2\sqrt{D_{\mbox{\scriptsize eff}}}$, discussed
in the previous section.
To check these ideas, we performed numerical simulations of
Eq.~(\ref{eq:rad}), with a FKPP production term (see appendix for
details on the numerical technique).  In Fig.~\ref{fig:2}, we show the
result of the calculations for the front speed $v_{\mbox{\scriptsize
f}}$ in dependence on $Da$, the slow reaction corresponds to the
plateau at $Da \ll 1$.

\subsection{Fast reaction regime}
\label{sec:3.3}
\begin{figure} 
\centerline{\includegraphics[scale=0.23,draft=false]{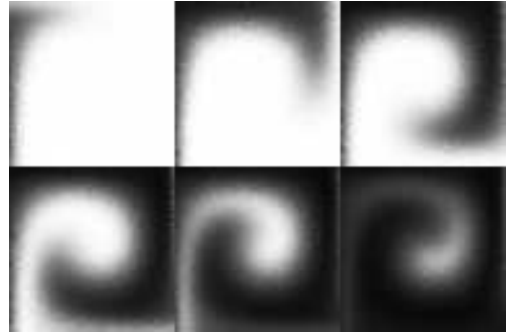}}
\vspace{0.4truecm}
\caption{Six snapshots of the field $\theta$ within the same cell,
at six successive times with a delay $(L/U)/6$ (left to right, top to bottom).
Here $Da=4,Pe=315$. A spiral wave invades the interior of the cell,
with a speed comparable to $U$.}
\label{fig:3}
\end{figure}

We have now to repeat the estimation of $D_{\mbox{\scriptsize eff}}$
and $\tau_{\mbox{\scriptsize eff}}$ for the fast reaction regime,
i.e., for large $Da$.  Since we work always in the regime of large
Peclet numbers, all the above arguments for the effective diffusion
still hold, while the effective time scale is different.  At large
$Da$, the ratio of time scales reverses, and in a (now short) time
$\tau$ two sharply separated phases emerge inside the cell.  In this
regime indeed, the interface is thin compared to the cell size. The
cell filling process is characterized by an inward spiral motion of
the outer, stable phase (see Fig.~\ref{fig:3}), at a speed
proportional to $U$, as it usually happens for a front in a shear flow
at large $Da$.  Therefore, the $\theta=1$ phase fills the whole cell
on the advective time scale, giving $\tau_{\mbox{\scriptsize
eff}}\simeq L/U$.
With respect to the upper bound (\ref{eq:upb}) we
observe for fast reaction a significative slowing-down of the front speed
signaled by a different scaling dependence on the parameters
$U,D,L,\tau$ (see Fig.~\ref{fig:2}).
\begin{figure} 
\centerline{\includegraphics[scale=0.33,draft=false,angle=270]{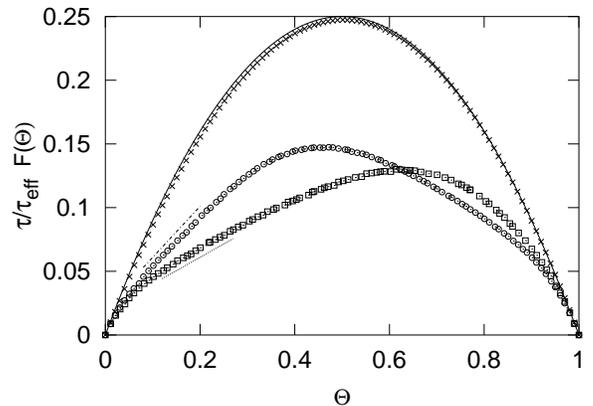}}
\caption{The renormalized reaction term 
$\tau / \tau_{\mbox{\scriptsize eff}}\, F(\Theta)$ for three different parameter:
$Da\simeq 4$ ($\Box$), $Da\simeq 2$ ($\circ$) 
and $Da \simeq 0.4 $ ($\times$).
The continuous line shows $f(\theta)$. 
The dotted and dash-dotted lines are the slopes ($0.2$ and $0.4$) 
proportional to $Da^{-1}$ 
in the region of slow advection.}
\label{fig:5}
\end{figure}
We have now to look at the shape of the effective reaction term
$\tau_{\mbox{\scriptsize eff}}^{-1} F(\Theta)$ appearing in the
renormalized equation~(\ref{eq:3}). As shown in Fig.~\ref{fig:5}, for
small $Da$, the effective production term is indistinguishable from
the ``bare'' one. Increasing $Da$, the reaction rate tends to reduce,
inducing the slowing-down of the front speed.  For $\Theta \approx 0$,
the effective production term essentially coincides with the
microscopic one. However, there is an intermediate regime
characterized by a linear dependence on the cell-averaged
concentration, with a slope proportional to $Da^{-1}$.  This is in
agreement with a typical effective reaction time
$\tau_{\mbox{\scriptsize eff}}\sim \tau Da$ (see below
Eq.~(\ref{eq:5})). To measure the macroscopic quantities $F(\Theta)$
and $\Theta$, one simply integrates numerically both $\theta$ and
$f(\theta)$ at a fixed time over a cell volume.

It is worth to remark that, notwithstanding the change of shape of the
effective chemical potential, the production term remains in the FKPP
universality class.

Let us stress that what we call fast reaction regime is still not in
the geometrical optics limit.  Indeed, to obtain this limit it is not
sufficient to take $\tau$ small, but one has to take also $D$ small in
such a way that $v_0 \sim 2\sqrt{D/\tau}$ is constant when $(D,\tau)$
goes to zero and the front thickness $\xi \propto \sqrt{D \tau}$ is
negligible with respect to the cell size. In the fast reaction regime
studied here the condition on the  front thickness holds.\\

\vspace{.4cm}

Collecting the information about fast and slow reaction
\begin{equation}
{\tau_{\mbox{\scriptsize eff}}\over \tau} \sim \left\{
\begin{array}{ll}
1  & \qquad Da \ll 1  \;\,\, \\
Da & \qquad Da \gg 1 \;.
\end{array}  \right.
\label{eq:5}
\end{equation}
and $D_{\mbox{\scriptsize eff}} \sim D Pe^{1/2}$, we can derive the scaling of the effective
speed of front propagation for a cellular flow.  Indeed, recalling that
$v_{\mbox{\scriptsize f}} \sim \sqrt{D_{\mbox{\scriptsize
eff}}/\tau_{\mbox{\scriptsize eff}}}$, we have the final result
\begin{equation}
{v_{\mbox{\scriptsize f}}\over v_0} \sim \left\{
{\begin{array}{lll}
Pe^{1/4} \qquad\,\qquad \;\;\;\; Da \ll 1 \,,\; Pe \gg 1 \;\,\, \\
Pe^{1/4}Da^{-1/2} \qquad Da \gg 1, Pe \gg 1 
\end{array}}  \right .
\label{eq:6}
\end{equation}
The case of $Pe \ll 1$ is less interesting because
the dynamics is dominated by diffusion.

\begin{figure} 
\centerline{\includegraphics[scale=0.33,draft=false,angle=270]{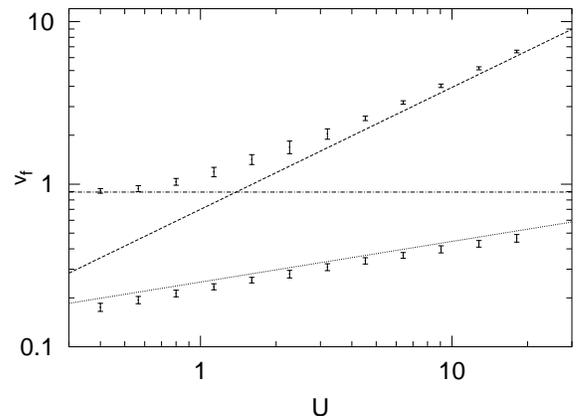}}
\caption{The front speed $v_{\mbox{\scriptsize f}}$ as a 
function of $U$, the typical flow velocity. The lower curve shows 
data at $\tau = 20.0$ (fast advection). 
The upper curve shows data at $\tau = 0.2$ (slow advection). For comparison, 
the scalings $U^{1/4}$ and $U^{3/4}$ are shown as dotted and dashed
lines, respectively. The horizontal line indicates $v_0$, 
the front velocity without advection, for $\tau = 0.2$.}
\label{fig:4}
\end{figure}

At small $Da$ the front propagates with an effective velocity scaling
as the upper bound derived above, that is as $Pe^{1/4}$.  At large
$Da$, the front speed enhanced is less effective than at small $Da$:
according to Eq.~(\ref{eq:6}), we have $v_{\mbox{\scriptsize
f}}/v_0 \sim Da^{-1/2}$ for $Da
\gg 1$.  In terms of the typical velocity of the cellular flow, we
have $v_{\mbox{\scriptsize f}} \propto U^{1/4}$ for slow reaction ($U
\gg L/\tau$, or equivalently $Da \ll 1$) whereas $v_{\mbox{\scriptsize
f}} \propto U^{3/4}$ for fast reaction ($U \ll L/\tau$, or $Da \gg
1$).  The scaling $v_{\mbox{\scriptsize f}} \propto U^{1/4}$ for slow
reaction (i.e. fast advection) is a consequence of the well known
result $D_{\mbox{\scriptsize eff}} \propto D {Pe}^{1/2}$ \cite{Pom} in
the homogenization limit~\cite{Const,abeletal}, it has been
obtained in~\cite{Pomeau}.  The numerical results are summarized in
Fig.~\ref{fig:4}.

As a remark we mention that, for the class of boundary conditions
investigated here, where the region of initially burnt material
extends to infinity, no quenching~\cite{quenching} takes place no
matter of the used production term.  Indeed, Arrhenius-type
nonlinearity substantially gives the same results as those of
FKPP-type reaction presented above, i.e.,  one has the two scaling laws
$v_{\mbox{\scriptsize f}} \propto U^{1/4}$ and $v_{\mbox{\scriptsize
f}} \propto U^{3/4}$ at fast and slow advection respectively (see
\cite{abeletal}).

\section{Front speed in the geometrical optics regime}
\label{sec:4}
When the front thickness and the reaction time are much smaller than
the length and time scales of the velocity field fluctuations one 
has the geometrical optics regime \cite{Majda-Geq,McLaughlin}.
In this case the front is a sharp interface separating the reactants
from the products, and can be modeled in the framework of the
$G$-equation~(\ref{eq:optics}) \cite{Peter,Majda-Geq}.
Physically speaking, one uses the $G$-equation when
the front thickness is very thin and it is hard
to resolve the diffusive scale.

As far as the cellular flow is concerned, the front border is wrinkled
by the velocity field during propagation and its length increases
until pockets of fresh material
develop~\cite{AS91,Aldredge94,Aldredge96}.  After this, the front
propagates periodically in space and time with an average speed
$v_{\mbox{\scriptsize f}}$, which is enhanced with respect to the
propagation speed $v_0$ of the fluid at rest~\cite{ottico} (see
Ref.~\cite{Aldredge94} for some pictorial views).

The problem addressed here is the dependence of the effective speed
$v_{\mbox{\scriptsize f}}$ on the flow intensity, $U$, and the bare
velocity, $v_0$, that is expected of the form \cite{Peter}:
\begin{equation}
{v_{\mbox{\scriptsize f}}\over v_0} =\psi\left({U \over v_0}\right)\,,
\end{equation}
where $\psi({\cal U}$) is a function
which depends on the flow details.

As far as we know, apart from very simple shear flows (for which
$\psi({\cal U}) = 1 + {\cal U}$ \cite{Pomeau,Majda-Geq}), 
there are no methods to compute
$\psi({\cal U})$ from first principles. Mainly one has to resort to numerical
simulations and phenomenological arguments.  

For turbulent flows, by means of dynamical
renormalization group techniques, Yakhot \cite{Yakhot88} proposed
\begin{equation}
{v_{\mbox{\scriptsize f}}\over v_0}= e^{(U/v_{\mbox{\scriptsize f}})^{\alpha}}
\label{eq:yakhot}
\end{equation}
with $\alpha=2$. Now $U$ indicates 
the root mean squared average velocity
(see also \cite{Peter,CY98}).
Therefore, from (\ref{eq:yakhot}) one has that  $v_{\mbox{\scriptsize f}}
\to  U/\sqrt{\ln(U)}$ for $U \to \infty $.  

\begin{figure}
\centerline{\includegraphics[scale=0.65,draft=false]{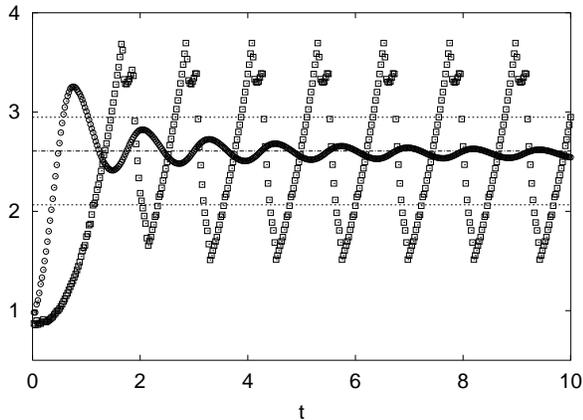}} 
\caption{Front velocity as a function of time, $v_{\mbox{\scriptsize
f}}(t)$ measured in the standard way (\ref{eq:velocity}) ($\Box$), and
as in Eq.~(\ref{eq:defvel2}) (o).  The straight lines represents the
average (over a period) of the measured value, and the lower and upper
bounds. The simulation parameters are $U=4$, $v_0=1$ and $L=2\pi$. }
\label{fig:ot2}
\end{figure}

\begin{figure}
\centerline{\includegraphics[scale=0.65,draft=false]{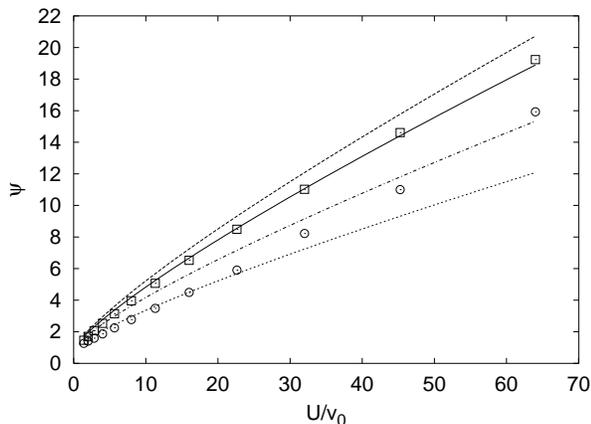}} 
\narrowtext
\caption{The measured $\psi(U/v_0)$ as a function of $U/v_0$ ($\Box$),
the Yakhot formula (\ref{eq:yakhot}) with $\alpha=2$ (o), the function
$\psi_{\beta}$ for $\beta=1,1/2$ (dashed and dotted lines) and for
$\beta=0.89$ (solid line).  The dashed-dotted line is the
bound~(\protect\ref{eq:oberman}).}
\label{fig:ot3}
\end{figure}

For the cellular flow under investigation, albeit the exact form of
the function $\psi({\cal U})$ is not known, a simple argument can be
given for an upper and a lower bound by mapping the front dynamics
onto a one-dimensional problem.  The starting point is the following
observation.  In the optical regime, since the interface is sharp,
i.e., $\theta(x,y)$ is a two-valued function ($\theta=1$ and
$\theta=0$), we can track the farther edge of the interface between
product and material ($x_{M}(t),y_{M}(t)$), which is defined as the
rightmost point (in the $x$-direction) for which
$\theta(x_M,y_M;t)=1$. Then we can define a velocity 
\begin{equation}
   \tilde{v}_f = \lim_{t \to \infty} {x_M(t) \over t}\,,
   \label{eq:defvel2}
\end{equation}
which gives an equivalent value of the standard definition 
within less than $2\%$ (see Fig.~\ref{fig:ot2}).
In Fig.~\ref{fig:ot2} we show the time evolution of the point
$(x_M(t),y_M(t))$.  After a transient, in the unit cell $[0,2\pi]$ (we
describe the case $L=2\pi$, i.e., the one adopted here) the point
$(x_M(t),y_M(t))$  moves to the right along the
separatrices of the streamfunction (\ref{eq:stream}), so that $y_M(t)$
is essentially close to the values $0$ or $\pi$.  Along this path one
can reduce the edge dynamics to the $1d$-problem
\begin{eqnarray}
{{\rm d}x_M \over {\rm d}t} = v_0 + U \beta |\sin(x_M(t))|
\label{eq:simplemodel}
\end{eqnarray}
where the second term of the r.h.s. is the horizontal component of the
velocity field. We have neglected the $y$-dependence, replacing it
with a constant $\beta$ which takes into account the average effect of
the vertical component of the velocity field along the path followed
by ($x_M,y_M$).  By solving (\ref{eq:simplemodel}) in the interval
$x_M\in (0,2\pi)$ one obtains the time, $T$, needed for $x_M$ to reach
the end of the cell.  The front speed, as the speed of the edge
particle, is then given by $v_{\mbox{\scriptsize f}} = 2\pi / T$. The
final result is 
\begin{equation}
\psi_{\beta}({\cal U}) = \pi {\sqrt{({\cal U} \beta)^2 - 1}}
\ln^{-1}\left( \frac{{\cal U}+\sqrt{({\cal U} \beta)^2 - 1}}
{{\cal U} -\sqrt{({\cal U} \beta)^2 - 1}}\right)\,,
\label{eq:func}
\end{equation}
Note that (\ref{eq:func}) is valid only for ${\cal U}\beta \geq 1$.
\begin{figure}
\centerline{\includegraphics[scale=0.65,draft=false]{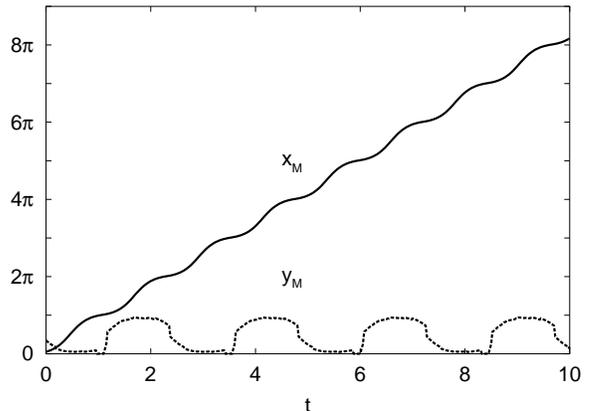}} 
\caption{Time evolution of the edge point:  $x_M(t)$ (solid line) and
$y_M(t)$ (dashed line). The simulation parameters are the ones of Fig.~\ref{fig:ot2}}
\label{fig:ot1}
\end{figure}
We have taken $\beta = 1$ for the upper bound and $\beta = 1/2$ (which
is the average of $|\cos(y)|$ between $0$ and $\pi$) for the lower
bound.  We have also computed the average of $|\cos(y_M(t))|$ in a
period of its evolution (see Fig.\ref{fig:ot2}) obtaining
$\beta\approx 0.89$ which gives indeed a very good approximation of
the measured curve (see Fig.~\ref{fig:ot3}).  We stress that the
theoretical curve is not a fit, but it just involves the measured
parameter $\beta$.

This agreement is an indication that the average of $|\cos(y_M(t))|$
depends on $U$ and $v_0$ very weakly (as we checked numerically).
Previous studies~\cite{Aldredge96} reported an essentially linear
dependence of the front speed on the flow intensity, i.e.,
$v_{\mbox{\scriptsize f}} \propto U$ for large $U$ which is not too far
but different from our result. A rigorous bound has been
obtained in Ref~\cite{oberman2001} by using the $G$-equation: 
\begin{equation}
v_{\mbox{\scriptsize f}} \geq U / (\log(1+U/v_0))\,\,.
\label{eq:oberman}
\end{equation}
As one can see from Fig.~\ref{fig:ot3}, the lower bound
(\ref{eq:oberman}) seems to be closer to the numerical data than the
one obtained with $\beta=1/2$ in (\ref{eq:func}).  From
Eq.~(\ref{eq:func}), asymptotically (i.e. for $U \gg v_0$) one has:
$v_{\mbox{\scriptsize f}} \sim U/\ln(U)$ which corresponds to
(\ref{eq:yakhot}) for $\alpha=1$. 
 Expressions as (\ref{eq:yakhot})
have been proposed for flows with many scales as, e.g., turbulent
flows, and in the literature different values of $\alpha$ have been
reported \cite{Peter,KA92}.  The fact that also the simple one-scale
vortical flow investigated here displays such a behavior may be
incidental.  However, we believe that it can be due to physical
reasons.  Indeed, the large scale features of the flow, e.g. the
absence of open channels (like for the shear flow) can be more
important than the detailed multi-scale properties of the flow
\cite{Ashurst}.

A definite answer to this question is beyond the scope of this paper,
however it could be an interesting point for future investigations.

\section{Conclusions}
\label{sec:5}
We addressed the problem of front speed enhancement induced by
stirring due to a cellular flow in different regimes.  In the slow
reaction case a rather general result (based on homogenization
techniques) gives $v_{\mbox{\scriptsize f}} \sim U^{1/4}$; for the
fast reaction case, physical arguments give $v_{\mbox{\scriptsize f}}
\sim U^{3/4}$.  In the geometrical optics limit one finds that
$v_{\mbox{\scriptsize f}}$ is a linear function of $U$, apart from
logarithmic corrections.  All these results has been confirmed by
numerical simulations.

The steady cellular flow treated here, albeit its simplicity,
provides a  paradigm that can be insightful for the
study of front propagation in more general flows. For instance, 
in the  geometrical optics limit the asymptotical
behavior  $v_{\mbox{\scriptsize f}} \sim U/\ln(U)$ 
is rather similar to the one found by Yakhot 
($v_{\mbox{\scriptsize f}}\sim U/\sqrt{\ln(U)}$)~\cite{Yakhot88} 
for turbulent velocity fields.  
\\
One could conjecture that there are nontrivial reasons for
this similarity: the front propagation principally involves
the largest scales. In this context, it is not surprising
that a multiscale process (as turbulence) or a single scale
process can yield similar results.

\section{Acknowledgments}
We gratefully thank A. Celani and A. Torcini, who recently
collaborated with us on the issue here discussed.  We also acknowledge
stimulating discussions with R. M. McLaughlin and
M. Vergassola.  This work has been partially supported by the INFM
{\it Parallel Computing Initiative} and MURST (Cofinanziamento {\it
Fisica Statistica e Teoria della Materia Condensata}).  M.A. has been
partially supported by the European Network {\it Intermittency in
Turbulent Systems} (FMRX-CT98-0175). M.C., D.V. and A.V. acknowledge
support from the INFM {\it Center for Statistical Mechanics and
Complexity} (SMC). 

\appendix

\section{Numerical method}
\label{app:1}
We introduce a lattice of mesh size $\Delta x$ and $\Delta y$ 
(for the sake of simplicity we assume $\Delta x = \Delta y$) 
so that the scalar field is defined on the points 
${\mathbf x}_{n,m} = (n\Delta x,m\Delta y)$: 
$\theta_{n,m}(t)=\theta(n\Delta x, m\Delta y, t)$.\\
Giving the field at time $t$, the algorithm computes
the field at time $t+\Delta t$, and the integration method
depends on which kind of physical regime we are interest in:
reaction-advection-diffusion equation (\ref{eq:rad}) or
optical limit (\ref{eq:optics}).

\subsection{Reaction Diffusion Equation}

In numerical approaches one is forced to discretize the dynamics, so
let us consider the case of a velocity field which is always zero apart
from $\delta$-impulses at times $t=0,\pm \Delta t, \pm 2\Delta t, \pm
3\Delta t, \ldots$
\begin{equation}
 {\mathbf u}({\mathbf x},t) = 
   \sum_{n=-\infty}^\infty {\mathbf u}({\mathbf x}) \delta(t - n\Delta t)\,\,.
 \label{eq:veldimp}
\end{equation}
In such a case the Lagrangian evolution is given by 
a conservative map (in 2d the map is symplectic due to incompressibility)
\begin{equation}
   {\mathbf x}(t+\Delta t) = {\mathbf F}^{\Delta t} ({\mathbf x}(t))\,\,.
   \label{eq:lagdifmap}
\end{equation}
If also the production term is zero apart from $\delta$-impulses 
\begin{equation}
   f(\theta) = \sum_{n=-\infty}^\infty g(\theta) \delta(t - n\Delta t)\,\,,
   \label{eq:reacdimp}
\end{equation}
one can introduce a reaction map
\begin{equation}
   \theta(t+\Delta t) = G_{\Delta t}(\theta(t))\,\,.
   \label{eq:reactmap}
\end{equation}
Let us remark that choosing a $\delta$-impulsed production term can be
particularly relevant in some experimental settings, i.e., when one
considers periodic illumination in light-sensitive chemical reactions
as in Ref.~\cite{strobo}.  The concentration field $\theta({\mathbf
x}, t + \Delta t - 0)$ is obtained from $\theta({\mathbf x}, t +
0)=G_{\Delta t}(\theta({\bf x},t))$ solving the bare diffusion
equation $\partial_t \theta = D \nabla^2 \theta$:
\begin{eqnarray}
\theta({\mathbf x}, t+\Delta t -0) = \frac{1}{(2\pi)^{d/2}} \nonumber
\\ \int e^{-\frac{\eta^2}{2}} \theta({\mathbf x} - \sqrt{2 D \Delta
t}\, {\boldmath {\mbox{$\eta$}}}, t + 0)\, {\mathrm d} {\boldmath
{\mbox{$\eta$}}}\,\,,
\label{eq:fwykacdimpmap}
\end{eqnarray}
or, in other terms:
\begin{equation}
   \theta({\mathbf x},t\!+\! \Delta t)\!=\!\left \langle\! { G_{\Delta t}(
          \theta({\mathbf F}^{-\Delta t}({\mathbf x}\!-\!\sqrt{2D
          \Delta t\,\,} {\boldmath {\mbox{$\eta$}}}(t)),t)) }\!\! \right
          \rangle_{\eta} \label{eq:feykacmap}
\end{equation}
which is equivalent to Eq.~(\ref{eq:feykacext}).
Let us remark that Eq.~(\ref{eq:feykacmap}) is exact if both the
velocity field and the reaction are $\delta$-impulsed processes.
However one can also use the formula (\ref{eq:feykacmap}) as a
practical method for the numerical integration of Eq.(\ref{eq:rad}) if
one assumes small enough $\Delta t$, so that the Lagrangian and reaction maps
are given at the lowest order by $$ {\mathbf F}^{\Delta t}({\mathbf x}) \simeq
{\mathbf x} + {\mathbf u}({\mathbf x}) \Delta t\,\,, \qquad G_{\Delta
t} \simeq \theta + \frac{\Delta t}{\tau_r} f(\theta)\,\,.$$

From an algorithmic point of view the whole process between $t$ and $t
+ \Delta t$, Eq.~(\ref{eq:feykacmap}), can be divided into three steps:
(1) diffusive, (2) advective and (3) reactive.  The first two steps
determine the origin of the Lagrangian trajectory evolving with a given
noise realization ${\boldmath {\mbox{$\eta$}}}$ and ending in ${\mathbf
x}$.  In the third step, the reaction at
point ${\mathbf x}$ for the advected/diffused passive scalar
$\theta$ is computed:
\begin{itemize}
   \item[1)] backward diffusion: 
             ${\mathbf x} \to {\mathbf x} - \sqrt{2D \Delta t}\,
             {\boldmath {\mbox{$\eta$}}}$
   \item[2)] backward advection by the Lagrangian map: \\
             ${\mathbf x} - \sqrt{2D \Delta t} \,{\boldmath {\mbox{$\eta$}}} \to 
             {\mathbf F}^{-\Delta t} ({\mathbf x} - \sqrt{2D \Delta t} 
             \, {\boldmath {\mbox{$\eta$}}})$
   \item[3)] forward reaction:\\
             $\theta(t+\Delta t) = G_{\Delta t}(\theta(t))$.
\end{itemize}
This three steps can be numerically implemented as follow.
For each grid point ${\mathbf x}_{n,m}$, 
one uses $N$ independent  Gaussian processes 
${\mathbf W}^\alpha$, $\alpha=1,\dots,N$, $N\gg 1$, and computes
${\mathbf {\tilde x}}^\alpha_{n,m}={\mathbf x}_{n,m} - 
  \sqrt{2D\,\Delta t}\, {\mathbf W}^\alpha$.
Then, using the Lagrangian backward propagator, 
${\mathbf r}^\alpha_{n,m} = 
 {\mathbf F}^{-\Delta t}({\mathbf \tilde{x}}_{n,m}^\alpha)$. 
For $\theta_{n,m}(t+\Delta t)$ one needs the values of $\theta$ at
time $t$ in the positions ${\mathbf r}_{n,m}^\alpha$. 
Typically the ${\mathbf r}_{n,m}^\alpha$ are not on the grid points 
$(n \Delta x, m \Delta y)$, nevertheless we can compute the value 
$\theta({\mathbf r}_{n,m}^\alpha,t)$ using simple linear interpolation 
from $\theta_{n,m}(t)$.
Therefore we have
$$\theta_{n,m}(t+\Delta t) = {1\over N} \sum_{\alpha=1}^N 
   G[\theta({\mathbf r}_{n,m}^\alpha,t)].$$
\begin{center}
\begin{figure}[htb]
\centerline{\includegraphics[scale=0.32,draft=false]{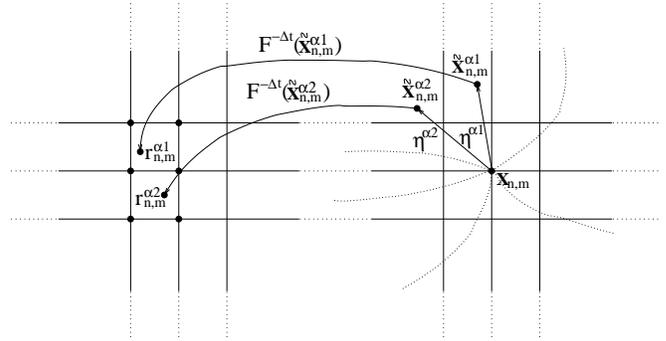}}
\narrowtext
\caption{Pictorial scheme of the numerical algorithm for RAD systems.
Here ${\boldmath {\mbox{$\eta$}}}^\alpha = \sqrt{2D\,\Delta t}\,
{\mathbf W}^\alpha$ where ${\mathbf W}^\alpha$ is a standard Gaussian
variable.}
\label{fig:nummet}
\end{figure}
\end{center}
To correctly simulate the diffusion process we have to impose a
relation between $D$, $\Delta x$ and $\Delta t$ to assure that
diffusion transports a particle over distances $\sim \sqrt{2D\,\Delta
t}$ much larger than the grid size $\Delta x $ (see
Fig~\ref{fig:nummet}).

\subsection{Geometrical optics limit}
Similar to the previous case, one can integrate the dynamics
of the optical front using a two step discrete-time process.
\begin{center}
\begin{figure}[htb]
\centerline{\includegraphics[scale=0.4,draft=false]{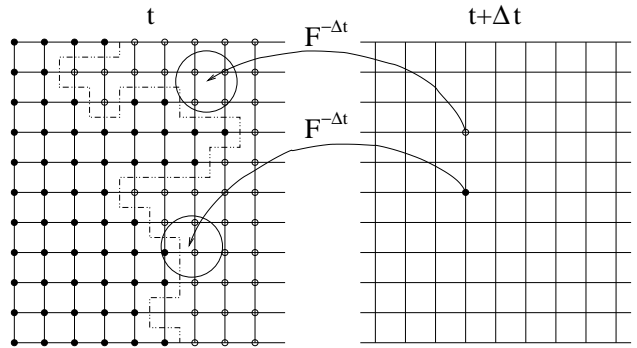}}
\narrowtext
\caption{Pictorial scheme of the numerical algorithm for the geometrical optics limit.}
\label{fig:nummetopt}
\end{figure}
\end{center}
Starting from the field $\theta_{n,m}$ at time $t$,
one can obtain the field at time $t+\Delta t$ with
the following algorithm
\begin{itemize}
   \item [1)] using the Lagrangian propagator one evolves the
   interface between burned and unburned region;
   \item [2)] at each point of the evolved interface one constructs
   a circle of radius $v_0\,{\rm \Delta}t$, burning the points
   within the circles.
\end{itemize}
To numerically implement such an algorithm one can proceed as follows:
starting in a grid point, ${\bf x}_{n,m}$, of the scalar field
at time $t + \Delta t$ one applies the backward evolution 
arriving at the point ${\bf y} = {\mathbf F}^{-\Delta t} {\bf x}_{n,m}$ 
at the time $t$. In this point we construct the circle of radius 
$v_0\,{\rm \Delta}t$. If in this circle there is at least
one burned point of the scalar field at time $t$, we fix 
$\theta({\bf x}_{n,m};t+{\rm \Delta}t) = 1$
otherwise $\theta({\bf x}_{n,m};t+{\rm \Delta}t) = 0$.

Also in this case we have to care about the radius of circle
$v_0\,{\rm \Delta}t$ has to be much larger than the grid size 
$\Delta x$.


\end{multicols}

\end{document}